\documentclass[prl,twocolumn,showpacs,superscriptaddress,nofootinbib]{revtex4-2}
\pdfoutput=1
\usepackage{graphicx}
\usepackage{epstopdf}
\usepackage{bm}
\usepackage{amssymb}
\usepackage{color}
\usepackage{amsmath}
\usepackage{amstext}
\usepackage{latexsym}
\usepackage[usenames,dvipsnames]{xcolor}
\usepackage[colorlinks=true,citecolor=MidnightBlue,linkcolor=RubineRed,urlcolor=MidnightBlue]{hyperref}

\def\x{{\rm\bf x}}

\newcommand{\beq}{\begin{equation}}
\newcommand{\eeq}{\end{equation}}
\newcommand{\beqa}{\begin{eqnarray}}
\newcommand{\eeqa}{\end{eqnarray}}

\usepackage{tikz,xcolor,hyperref}
\definecolor{lime}{HTML}{A6CE39}
\DeclareRobustCommand{\orcidicon}{
\begin{tikzpicture}
\draw[lime, fill=lime] (0,0)
circle[radius=0.16]
node[white]{{\fontfamily{qag}\selectfont \tiny \.{I}D}}; 
\end{tikzpicture}
\hspace{-2mm}
}
\foreach \x in {A, ..., Z}{%
\expandafter\xdef\csname orcid\x\endcsname{\noexpand\href{https://orcid.org/\csname orcidauthor\x\endcsname}{\noexpand\orcidicon}}
}

\begin{document}

\title{Spontaneous Symmetry Breaking of Vortex Number in Binary Alternating Current Countersuperflow}

\author{Wei-Can Yang\hspace{-1.5mm}\orcidA{}
}
\affiliation{Department of Physics, Osaka Metropolitan University, 3-3-138 Sugimoto, 558-8585 Osaka, Japan}

\author{Makoto Tsubota}\email{tsubota@omu.ac.jp}
\affiliation{Department of Physics, Osaka Metropolitan University, 3-3-138 Sugimoto, 558-8585 Osaka, Japan}
\affiliation{Nambu Yoichiro Institute of Theoretical and Experimental Physics (NITEP), Osaka Metropolitan University, 3-3-138 Sugimoto, Sumiyoshi-ku, Osaka 558-8585, Japan}

\author{Hua-Bi Zeng\hspace{-1.5mm}\orcidB{}
}\email{hbzeng@yzu.edu.cn}
\affiliation{Center for Theoretical Physics , Hainan University, Haikou 570228, China}
\affiliation{Center for Gravitation and Cosmology, College of Physical Science and Technology, Yangzhou University, Yangzhou 225009, China}

\begin{abstract}
In binary superfluid counterflow systems, vortex nucleation arises as a consequence of hydrodynamic instabilities when the coupling coefficient and counterflow velocity exceed the critical value. When dealing with two  identical  components, one might naturally anticipate that the number of vortices generated would remain equal. However, through the numerical experiments of the holographic model and the Gross-Pitaevskii equation, our investigation has unveiled a remarkable phenomenon: in Alternating Current counterflow systems, once the coupling coefficient and frequency exceed certain critical values, a surprising symmetry-breaking phenomenon occurs. This results in an asymmetry in the number of vortices in the two components. We establish that this phenomenon represents a novel continuous phase transition, which, as indicated by the phase diagram, is exclusively observable in Alternating Current counterflow. We provide an explanation for this intriguing phenomenon through soliton structures, thereby uncovering the complex and unique characteristics of quantum fluid instabilities and their rich phenomena.
\end{abstract}

\maketitle

The complexity of fluid dynamics stems from its inherent instability, providing valuable insights into a wide range of phenomena, including pattern formation \cite{Cross_1993,Imperio_2004,L_cke}, vortex and soliton emergence \cite{Grosso_2011,Amo_2011,Francois_2014}, and turbulence \cite{Koop_1979,Thorpe_1987,Swinney_1978}. These instabilities bridge the dynamics of classical and quantum fluids, revealing commonalities\cite{Kobayashi_2005,Chen_2006,Reeves_2013}, with examples like Kelvin-Helmholtz \cite{Takeuchi_2010, FUNADA_2001} and Rayleigh-Taylor instabilities \cite{Vinningland_2007, Sasaki_2009} seen in both.  However, unlike the complexity of classical fluids, quantum fluids are characterized by the presence of ideal quantized vortices \cite{barenghi2023quantum,Tsubota_2013}, offering a unique opportunity to study turbulent dynamics in a primitive environment.  Thus, they provide a key platform for exploring the nexus of instabilities and complex flow phenomena.

In quantum fluids, such as superfluid helium and Bose-Einstein Condensates, a significant consequence of instability manifests as the emergence of quantum turbulence \cite{Tsubota_2013,Bewley_2006,Salomaa_1987}. 
This phenomenon has garnered substantial attention as an idealized prototype of classical turbulence.  Quantum turbulence is generated primarily through two counterflow methods: thermal counterflow involving normal and superfluids \cite{Schwarz_1978,Guo_2010,lan2020landau,PhysRevB.87.174508}, and countersuperflow between two superfluids, causing turbulence in both \cite{PhysRevLett.90.100401,Hamner_2011,Hoefer_2011,Takeuchi_2010_2,Ishino_2011}.
Our primary emphasis is directed towards the latter counterflow scenario due to its readily controllable attributes, which yield a wealth of observable phenomena.  Experimentally, countersuperflow instability is induced in two-component Bose-Einstein Condensates using magnetic-field gradients, leading to opposing velocities \cite{Hamner_2011,Hoefer_2011}.  This instability in quasi-one-dimensional systems initiates dark-bright solitons, which in higher dimensions nucleate quantized vortices, crucial for quantum turbulence \cite{Takeuchi_2010_2}.

Numerical and theoretical studies on countersuperflow instability reveal similar findings \cite{Ishino_2011}. Furthermore, Linear instability analysis using the Gross-Pitaevskii equation and Bogoliubov-de Gennes model helps precisely define instability conditions \cite{Takeuchi_2010_2}.  By analyzing the Bogoliubov energy spectrum's dispersion relation, we can determine the requirements for momentum exchange and instability leading to excitation:
\begin{eqnarray}
    \delta J_1 &=& \delta J_2 \neq 0 \nonumber \\
    U_{12} &>& U_{-}
    \label{Instability}
\end{eqnarray}
Here, in the context of momentum conservation, $J_{i}$ represents the momentum of component $i$, and $\delta J_{i}$ is the exchanged momentum. Momentum exchange between components must be equal. Instability and subsequent excitation occur when the relative velocity $U_{12} = |{\bf U_1} - {\bf U_2}|$ exceeds the critical velocity $U_-$.

Previous investigations of countersuperflow instability have primarily focused on Direct Current (DC) counterflow, where vortices fully nucleate due to evolution over a sufficiently long period of time to present an equal number of vortices in two components \cite{Takeuchi_2010_2}. However, when the analysis shifts to Alternating Current (AC) counterflow, the complexity increases significantly due to the system's immersion in continuous excitation processes. As a result, fully understanding these systems may require a more holistic approach that goes beyond just examining the statistical behavior of vortices.

\begin{figure}[t]
    \centering
    \includegraphics[width=11.5cm,trim=60 30 -80 0]{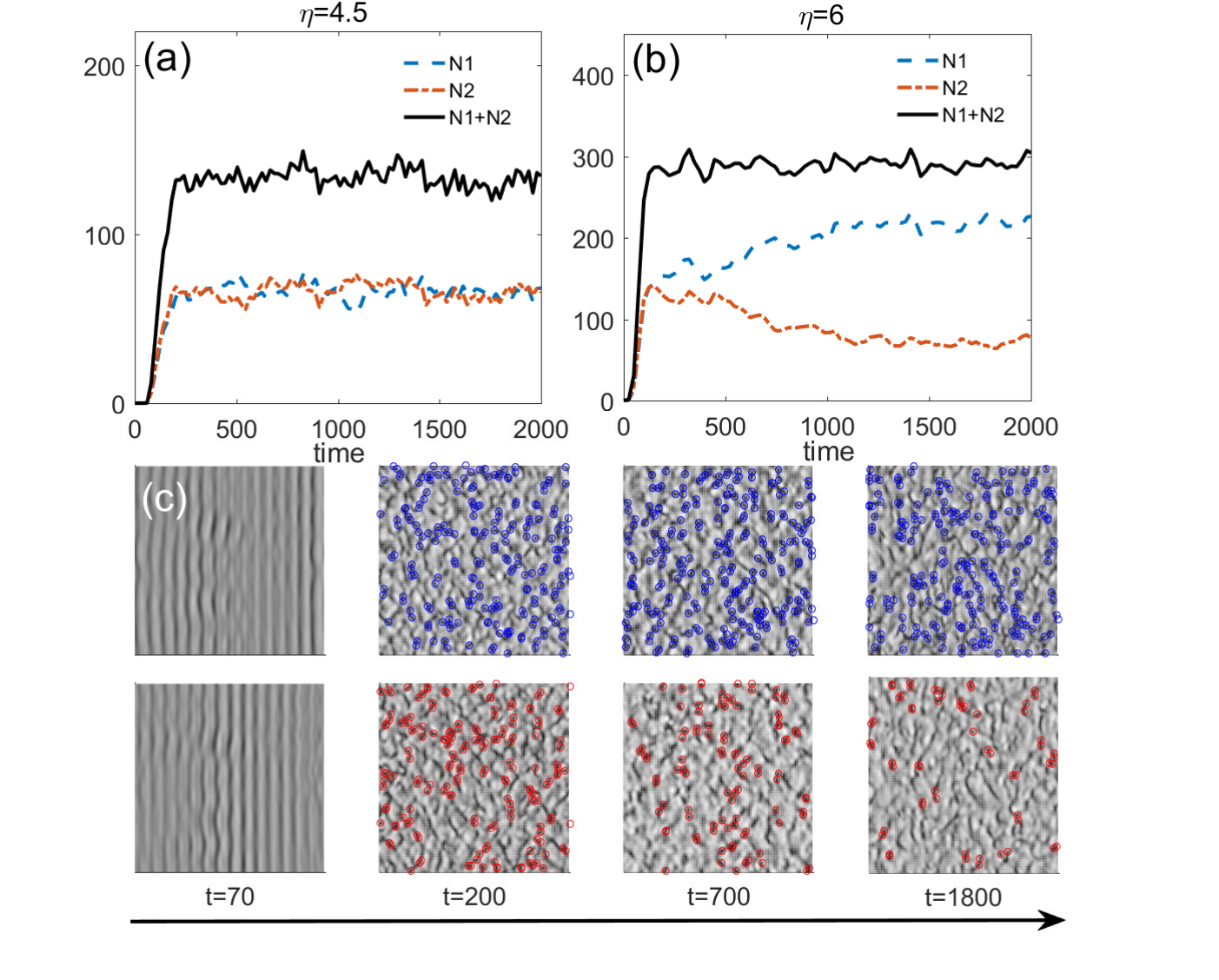}
    \caption{{\bf (a):} The vortex evolution with coupling constant $\eta=4.5$, vortex number exhibits completely symmetrical behavior. {\bf (b):} The vortex evolution with coupling constant $\eta=6$, symmetry breaking occurs in the number of vortices.
    {\bf (c):} The dynamic evolution diagram of the densities corresponding to Figure (b). The upper and lower rows refer to components $\Psi_1$ and $\Psi_2$.}
    \label{figure1}
\end{figure}

In our simulation, we initially employ the holographic superfluid model. This model utilizes the duality between high-dimensional gravitational fields and low-dimensional gauge fields to explore superfluid dynamics at the boundary \cite{PhysRevD.79.066002}. This approach has been widely used in the study of superfluid  vortices and quantum turbulence, especially for strongly correlated systems \cite{Chesler_2013,Lan_2016,PhysRevB.107.144511}, enables us to make meaningful comparisons with real-world physical systems \cite{PhysRevLett.127.101601,Du_2015}.
Here, we adopt a bottom-up holographic model, where $(2+1)$-dimensional superfluid behavior is characterized by an Abelian-Higgs model existing in a $(3+1)$-dimensional asymptotically anti-de Sitter (AdS) black hole spacetime\cite{PhysRevB.107.144511}. The action is formulated as follows:
\begin{eqnarray}
S &= \int d^4x \sqrt{-g} \Big[-\sum_{j=1}^2(\frac{1}{4}F^2_j-|D_j\Psi_j|^2-m_j^2|\Psi_j|^2)+V\Big] \nonumber \\
 &V(\Psi_1,\Psi_2) = \eta|\Psi_1|^2|\Psi_2|^2
\end{eqnarray}
Here $F_{j}=\partial_\alpha A_{j\beta}-\partial_\beta A_{j\alpha}$ represents the Maxwell field strength with vector potential $A_{j\alpha}$, which is coupled minimally to the scalar involving the charge $q$ with the covariant derivative $D_j=\partial_{j\alpha}-iqA_{j\alpha}$. The complex scalar field $\Psi_j$ has mass  $m_j$ and $\eta$ is the inter-component coupling constant.

When the chemical potential $A_t|_{z=0}=\mu$ exceeds the critical value $\mu_c=4.07$, the system cools down into the superfluid phase. We prepare two identical and static initial superfluid components by setting $\mu_1=\mu_2=9$.
To investigate AC counterflow, we activate the spatial gauge field component $A_{jx}$  to induce completely opposite and oscillating flow fields in the two components:
\begin{eqnarray}
    v_{jx}(t)=A_{jx}(t,z=0)=(-1)^{j-1}U\cos(\Omega t)
    \label{velocity}
\end{eqnarray}
Here $v_{jx}(t)$ represents the velocity of $j$-th component. $U$ stands for the amplitude, while $\Omega$ denotes the frequency of the applied velocity field. It is worth noting that we set $U=3$, which  exceeds half of the critical counterflow velocity $U_{-}/2$ when $\eta$ is less than the phase separation critical value $\eta_{ps}=7.05$ \cite{Yang_2020}.

\begin{figure}[t]
    \centering
    \includegraphics[width=9cm, trim= 10 30 0 0]{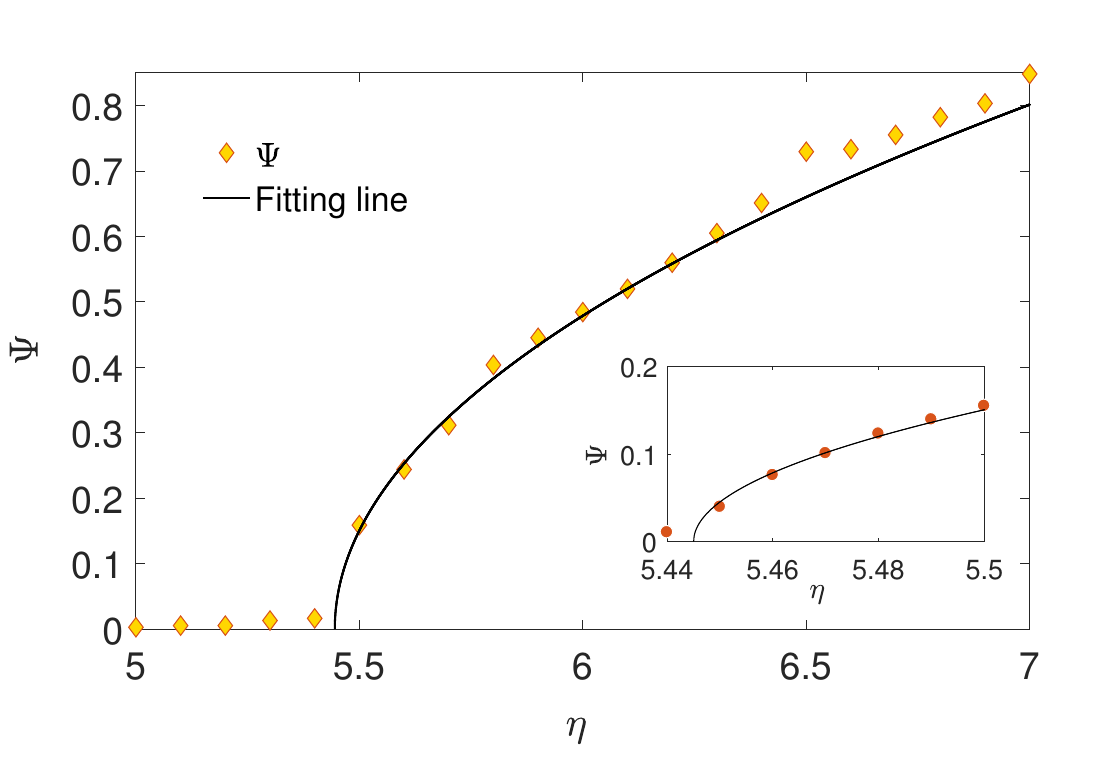}
    \caption{Order parameter $\Psi$ as a function of $\eta$ for $\Omega=0.1\mu_c$, $\Psi=(1.542\pm 0.11)(\eta/5.445-1)^{0.507\pm0.027}$.   The black line is the fit curve, and inset magnifies the critical region. The spacing of the selected points in the large figure is $\Delta \eta =0.1$, while in the inset figure is $\Delta\eta=0.01$.}
    \label{figure2}
\end{figure}

 Anticipated are that, owing to the identical characteristics of the two superfluid components and their symmetrical positioning in the Lagrangian, the number of vortices generated by counterflow should also persist in equality\cite{Supplementary1}.
Fig.\ref{figure1}a and \ref{figure1}b depict the evolution of vortices in the counterflow under two distinct coupling conditions $\eta=4.5$ and $\eta=6$ with  $\Omega=0.1\mu_c$. The number of vortices presented in the figure has undergone temporal averaging, resulting in a smoother and more easily interpretable representation that eliminates oscillations. In cases characterized by small coupling coefficients, as anticipated, the vortex counts for both components consistently remain equal, with no special behaviors emerging. 
However, under conditions of large coupling, the vortex counts in the two components gradually diverge, ultimately reaching a non-equilibrium steady state characterized by a stable difference in the number of vortices, denoted as $|N_1-N_2|$. We have verified that it is completely random which one of the two components will have the dominant number of vortices.
It is noteworthy that the total vortex count, denoted as $N=N_1+N_2$, remains constant despite the deviation in the number of vortices. 
This constitutes a novel and intriguing phenomenon, wherein asymmetric outcomes arise from perfectly symmetric systems. We posit that this represents a newfound occurrence of symmetry breaking in vortex statistical systems. Fig. \ref{figure1}c  shows the dynamic process of vortex number symmetry breaking, and the related movie can be found in the supplementary \cite{Supplementary2}.

At a fixed frequency of $\Omega=0.1\mu_c$, we conduct an exhaustive exploration of the symmetry breaking as a function of the coupling coefficient. Here, the order parameter is defined as the ratio of the difference in vortex numbers to their sum:
\begin{equation}
   \Psi(\eta,\Omega)=|N_1-N_2|/N 
\end{equation}

As observed in Fig. \ref{figure2}, the order parameter is entirely continuous. 
The fitting results indicate that in the vicinity of the critical point, the variations in the order parameters conform to a mean field phase transition, with the critical exponent $\nu=0.5$.

\begin{figure}[t]
    \centering
    \includegraphics[width=9cm,trim= 10 10 0 0]{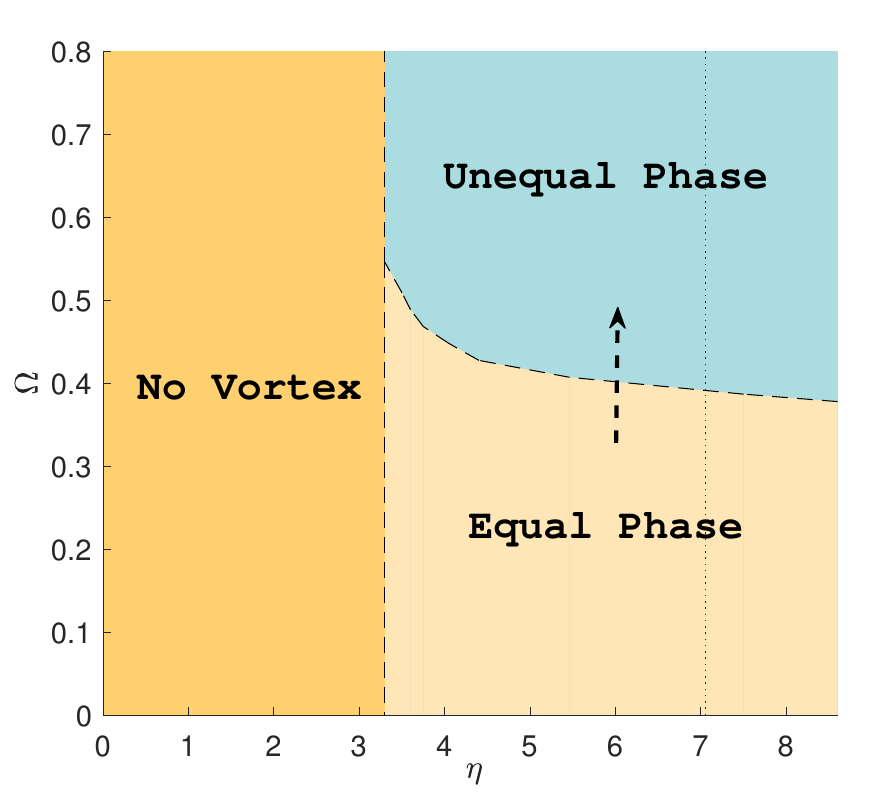}
    \caption{The comprehensive phase diagram, delineated by oscillation frequency $\Omega$ and coupling coefficient $\eta$, where the sampling  points are spaced $0.1$, be partitioned into three distinct regions: the left phase represents the absence of vortex, the lower-right phase signifies a phase with symmetric vortex numbers, and the upper-right phase designates a phase with broken vortex symmetry. 
    The dot line on $\eta_{ps}=7.05$ is the critical boundary between miscible and immiscible phase, above which the two components are  phase separated.}
    \label{figure3}
\end{figure}

\begin{figure}[h]
    \centering
    \includegraphics[width=9.5cm,trim=40 30 0 0]{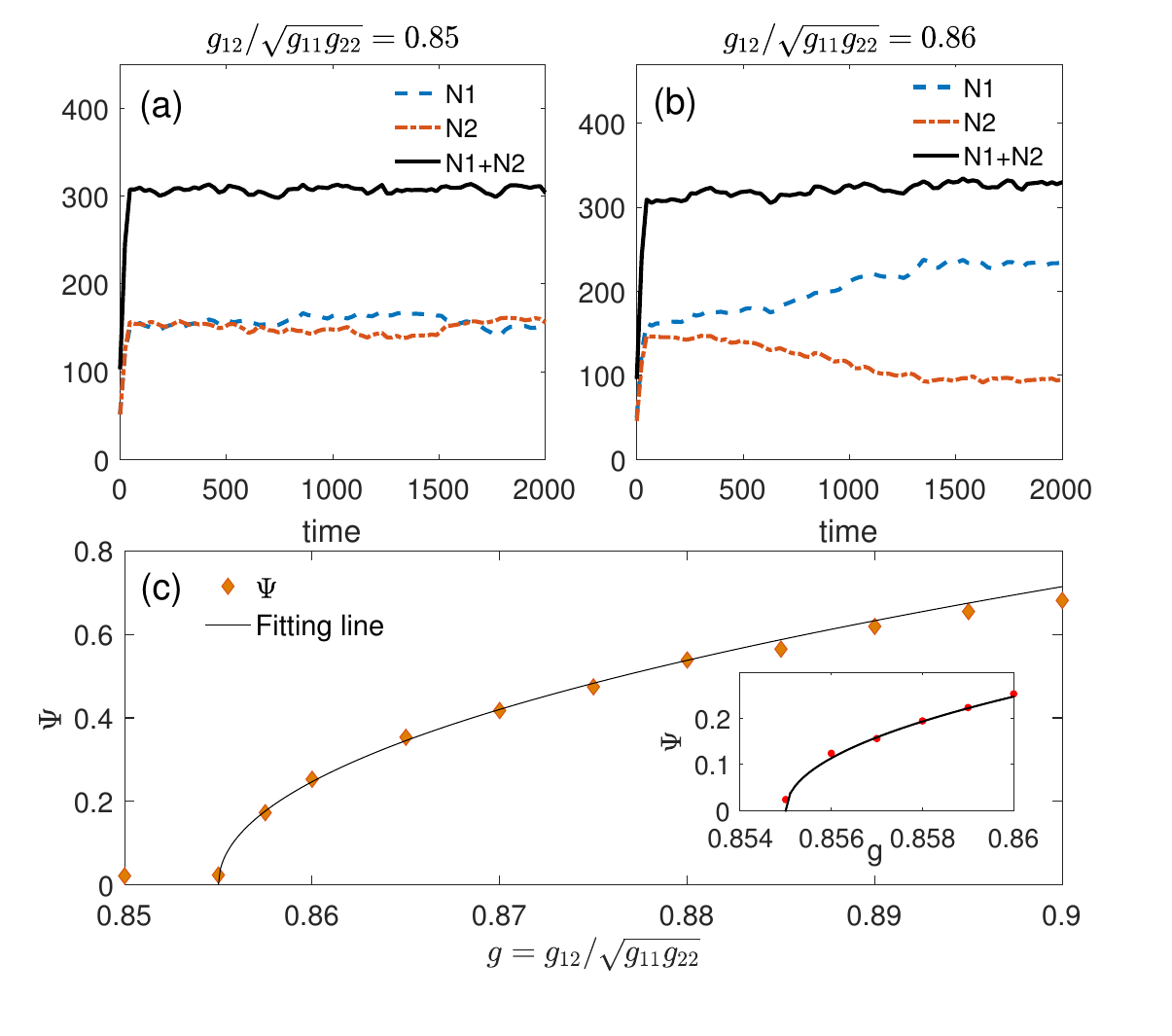}
    \caption{The evolution of vortex numbers in Gross-Pitaevskii equation, {\bf (a):} The vortex number maintains symmetry. {\bf (b):} The vortex number exhibits symmetry breaking.
    {\bf (c):} Order parameter $\Psi$ as a function of coupling coefficient $g_{12}$ for $\Omega=0.4$, $\Psi=(2.957\pm0.383)(g/0.855-1)^{0.475\pm0.031}$. The black line is the fit curve, and the small graph at the bottom right magnifies the critical region. }
    \label{figure4}
\end{figure}

Furthermore, our observations reveal that the symmetry breaking of vortex numbers is also influenced by the oscillation frequency $\Omega$. There exists a critical frequency $\Omega_c$ as a function of $\eta$, above which a breakdown of vortex number symmetry occurs. Conversely, below this critical frequency, the vortex numbers remain equal. Fig. \ref{figure3} 
presents the complete phase diagram encompassing both the coupling coefficient and the oscillation frequency. The left region of the diagram corresponds to scenarios where the coupling coefficient is too low $\eta<\eta_v=3.2$ to nucleate vortices. It is noteworthy to mention that, since the vortex number is an average over time, its definition becomes less precise in the low-frequency region. Nevertheless, this does not substantially impact the overall discussion of the phase diagram. When $\eta$ exceeds the critical value, the total vortex number increases linearly with coupling like $N=k(\eta-\eta_v)$, see supplementary \cite{Supplementary1}.
In the lower-right region, where neither the oscillation frequency nor the coupling coefficient is sufficiently high, complete symmetry is maintained with equal numbers of vortices. As the oscillation frequency and coupling coefficient gradually increase, the system traverses the critical curve, leading to symmetry breaking. This transition ushers the system into a phase characterized by differing vortex numbers. It is important to emphasize that even with a sufficiently large coupling coefficient, the symmetry breaking of vortex numbers necessitates a finite oscillation frequency. As a result, this symmetry-breaking behavior cannot be observed in DC counterflow, where DC counterflow corresponds to the case of $\Omega=0$.

As a comparison and generalization supplement, we perform the same simulation for the Gross-Pitaevskii model. Unlike the holographic model, which is suitable for describing strong correlation and strong dissipation, the Gross-Pitaevskii equation is  powerful for describing weakly correlated systems.In two dimensions, the two-component Gross-Pitaevskii equation can be simply written as
\begin{equation}
    i\hbar\frac{\partial}{\partial t}\Psi_j = (-\frac{\hbar^2}{2m_j}\nabla^2+\sum_kg_{jk}|\Psi_k|^2-{\bf u}_j \cdot {\bf p})\Psi_j
\end{equation}
To maintain generality and ensure symmetry between the two components, we use dimensionless parameters: $\hbar=m_1=m_2=1$ and $g_{11}=g_{22}$.
The external flow is set by ${\bf u}$\cite{Reeves_2015}. We set a counterflow condition of periodic oscillation
$v_{jx}(t)=(-1)^{j-1}U \cos (\Omega t)$. As shown in Fig.\ref{figure4}, the simulation of the Gross-pitaevskii equation shows similar results to holographic model.
With minimal coupling, the number of vortices remains symmetrical; however, increased coupling leads to symmetry breakdown.
Fig.\ref{figure2} and Fig.\ref{figure4}c show a qualitative consistency in the order parameters of the two models, a consistency that is also validated in the vortex lattice phase diagram\cite{Tsubota2003,Yang_2020}, where $g=1$ in the Gross-Pitaevskii equation corresponds to $\eta=7$ in the holographic model. However, due to intrinsic differences between the holographic model and the Gross-Pitaevskii model, they cannot be quantitatively consistent.

For more profound understanding of vortex number symmetry breaking, we investigate the actual velocity field \cite{Kobayashi_2005}. The order parameter $\Psi=\sqrt{n} \exp(i\phi)$ can be separated into its superfluid density $n$ and phase $\phi$. The actual velocity of the superfluid can be calculated by the gradient of the phase $\nabla\phi$.
\begin{eqnarray}
    {\bf v}=\frac{i[\Psi^{*}\nabla\Psi-\Psi\nabla\Psi^{*}]}{2|\Psi|^2}
\end{eqnarray}
We decompose the velocity vector into its respective $x$ and $y$ components and average each component over the full space;
\begin{eqnarray}
   \langle v_{x,y} \rangle = \frac{\int v_{x,y} ds}{\int ds} 
\end{eqnarray}

\begin{figure}[t]
    \centering
    \includegraphics[width=10cm,trim=30 0 0 0]{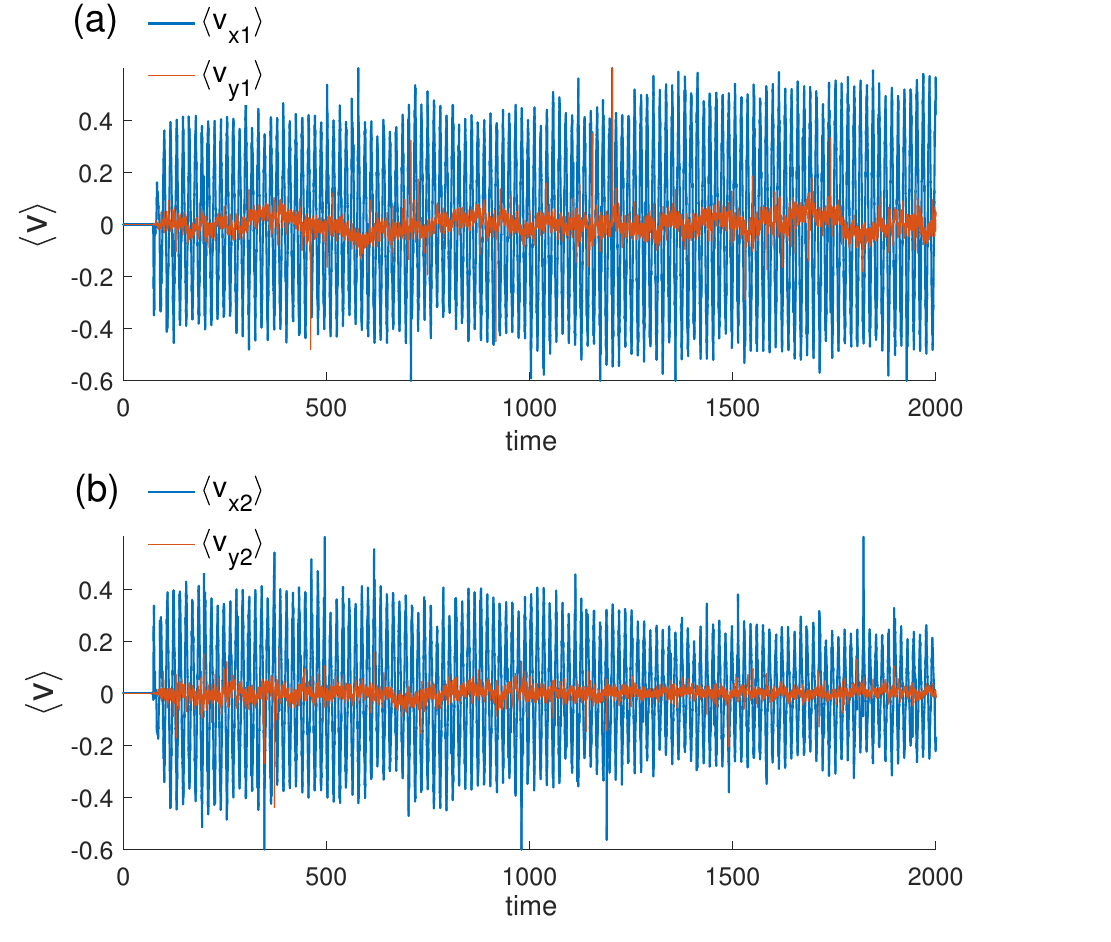}
    \caption{In the vortex number unequal phase with $\eta=6$, $\Omega=0.1\mu_c$. In both figures denoted as {\bf (a)} and {\bf (b)}, the red line symbolizes the averaged velocity in the $y$-direction, whereas the blue line represents the averaged velocity in the $x$-direction.
In component 1 of figure(a), the peak value or amplitude of $\langle v_x\rangle$ increases with the increase in vortex number, while in component 2 of figure(b), the peak value or amplitude of $\langle v_x\rangle$ decreases with the decrease in voertx number.}
    \label{figure5}
\end{figure}

In Fig. \ref{figure5}, we show the results of the average velocity of the two components in the $x$ and $y$ directions in the case of vortex number symmetry breaking with parameter $\eta=6$, $\Omega=0.1\mu_c$.
Due to the oscillation of the system's background flow, the overall velocities oscillate too. Notably, the background velocity we have introduced as Eq.(\ref{velocity}) operates exclusively in the $x$-direction, resulting in an overall $y$-direction velocity of zero. 
Importantly, the amplitude of $\langle v_y\rangle$ remains constant regardless of the generation or number of vortices, indicating that vortex count has no bearing on the $\langle v_y \rangle$.
Conversely, in the x-direction, the peak value or amplitude of $\langle v_x \rangle$ varies over time. More precisely, the component with a growing vortex count encounters an escalation in amplitude (Fig. \ref{figure5}a), whereas the component with a diminishing vortex count exhibits a reduction in amplitude (Fig. \ref{figure5}b). This observation, in conjunction with the notion that vortex count exerts no influence on the mean velocity, strongly implies the existence of an additional structural element, separate from vortices, which affects the overall velocity.

\begin{figure}[t]
    \centering
    \includegraphics[width=10.5cm,trim=30 0 0 0]{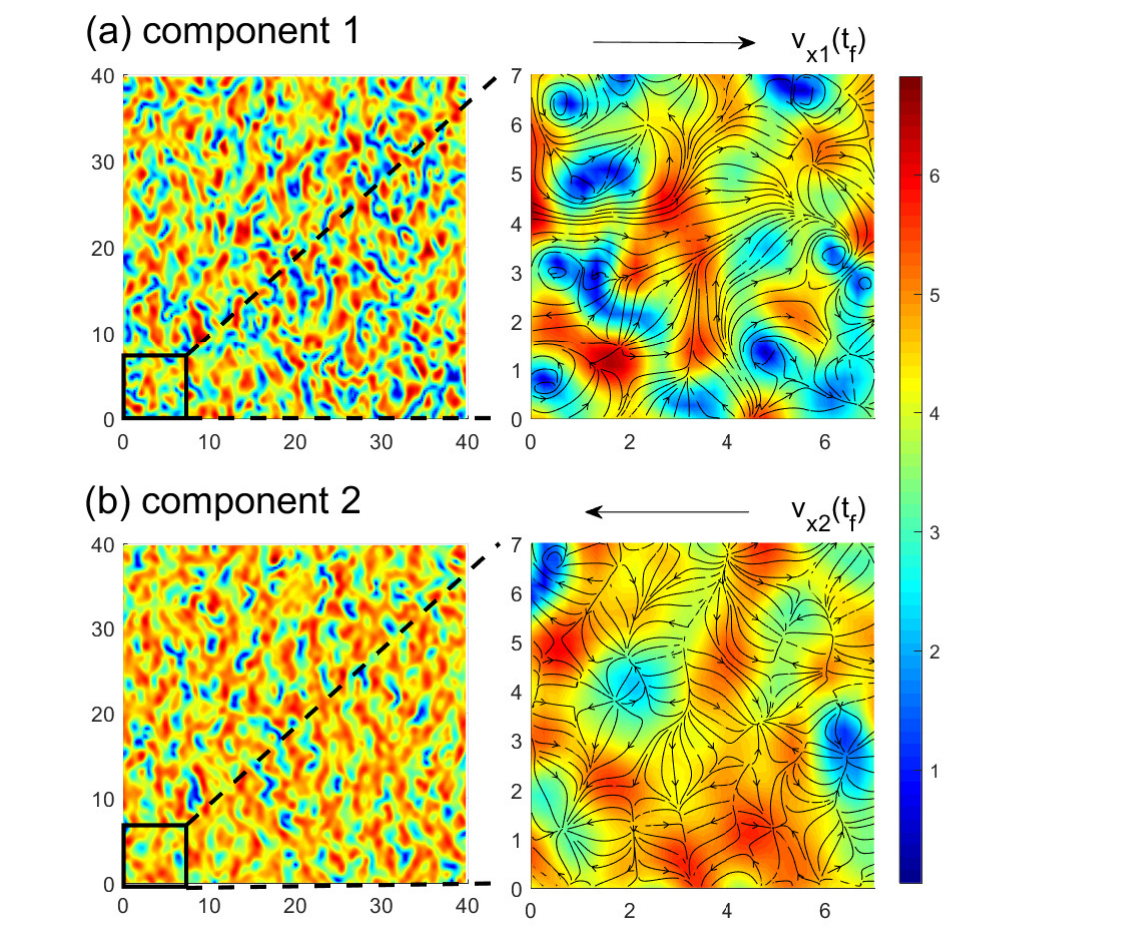}
    \caption{At $t=2000$ where vortex number symmetry is completely broken. {\bf (a)} and {\bf (b)} panel shows the density profile and velocity field of component 1 and component 2, respectively. In both panel, left figure shows the complete diagram while right figure shows the local enlarged diagram. }
    \label{figure6}
\end{figure}

Additional evidence can be discerned by magnifying the details in the flow velocity field distribution diagram, as depicted in Fig.\ref{figure6} at $t=2000$. 
In this case, Component 1 has a greater number of vortices, while Component 2 has fewer vortices. The visual representation encompasses the full image on the left, with the right portion providing an enlarged perspective.
At this juncture, the background flow velocity for Component 1 is directed from left to right, while for Component 2, it flows in the opposite direction. Upon meticulous examination of the locally magnified view in Component 1, it becomes apparent that, despite the presence of numerous vortices, the overall velocity remains congruent with the background velocity. This observation aligns with prior conclusions that vortices do not exert a significant impact on the overall velocity.
Conversely, in the locally magnified view of Component 2, the flow velocity does not conform to the background flow velocity, despite the absence of vortices. Indeed, across numerous regions, the velocity counters the background veloicty. This compelling observation provides strong evidence that the reduction in the amplitude of $\langle v_{x2}\rangle$ is attributed not to vortices but rather to a distintive structural element known as the soliton structure. This soliton structure exhibits a flow pattern that opposes the background flow. A more detailed and clear diagram of soliton structures is disscussed in the supplementary \cite{Supplementary1}.

Incorporating the flow velocity field with the phase diagram illustrating vortex symmetry breaking, we can conclude that vortex symmetry breaking arises from the existence of soliton structures that do not undergo vortex nucleation. 
To generate a vortex pair, a soliton requires a characteristic time $t_{soliton}$. When the characteristic time, as indicated by the oscillation frequency $\Omega$, is shorter than the average $t_{soliton}$, the solitons may not fully decay, resulting in the absence of local couterflow when the flow velocity reverses: $\hat{t}=\frac{\pi}{\Omega} < \langle t_{soliton} \rangle$.

Simultaneously, since solitons contribute to the momentum exchange, momentum conservation as described in Eq.(\ref{Instability}) encompasses not only vortex momentum but also soliton momentum.  Therefore, the number of vortices no longer has to remain symmetric, and the instability condition becomes
\begin{eqnarray}
    \delta J_{vortex1}+\delta J_{soliton1} = \delta J_{vortex2}+\delta J_{soliton2}
\end{eqnarray}
Throughout the process of vortex number symmetry breaking, 
the number of solitons decreases in the component with more vortices, in contrast to the increase in soliton count in the component with fewer vortices, so that the average velocity changes. Owing to pronounced non-linearity, the attainment of absolute symmetry between the two components at the outset is unrealistic, inevitably leading to minor deviations. Based on our validation, as shown in the supplementary materials \cite{Supplementary1}, the component that displays higher velocity in the counterflow is inclined to generate voritces. 
In contrast, solitons within the component characterized by lower velocities experineces a more gradual and attenuated process of splitting, leading to a decreased vortices count. Therefore, when solitons are present and their number varies, the velocity deviation gradually amplifies.
Consequently, these minor deviations gradually amplify, leading to more pronounced symmetry breaking. The final non-equilibrium steady state emerges from a balance between the number of solitons and the rate at which they split. This complex interplay ultimately defines the dynamic behavior of the system in its non-equilibrium state.

Our research on AC counterflow uncovers a novel dynamic phase transition, induced when oscillation frequency and coupling surpass critical values, leading to asymmetry in vortex numbers as predicted by mean field theory. The phase diagram explains the absence of this symmetry breaking in DC counterflow. Analysis of flow velocities reveals that this transition stems from incompletely evolved soliton structures within each oscillation period. These solitons, participating in momentum exchange, disrupt the expected vortex number symmetry and alter the mean flow velocity, intensifying the asymmetry. Supplementary material \cite{Supplementary1} includes additional cases, like beyond-mean-field models and non-periodic systems, supporting our findings' universality.
Experimentally, we expect that AC counterflow can be achieved through the Zeeman shift driven by periodic magnetic field gradients \cite{Hamner_2011}, allowing us to explore the breaking of vortex number symmetry in practical applications.

{\it Acknowledgements.}
H.B. Z. acknowledges the support by the National Natural Science Foundation of China (under Grants No. 12275233), M. T. acknowledges the support by the
JSPS KAKENHI (under Grant No. JP20H01855).

\onecolumngrid
\appendix 
\clearpage
\section*{Supplementary Material}

 In the supplementary material, we begin with a comprehensive introduction to the holographic superfluid model.  Following this, we delve into the results obtained from simulation of beyond-mean field system and non-periodic system, thereby highlighting the universality of the observed phenomenon across different models.
We then present a detailed visual representation of the soliton structure. This visual aid is instrumental in describing the causes behind the observed variations in velocity within the soliton. Additionally, through a comparison of equal-velocity and non-equal-velocity Direct Current (DC) counterflow, we further explore and clarify the mechanisms underlying the symmetry breaking observed in these systems.
After that, we show that how the total number of vortices varies in relation to the coupling coefficient and how phase diagram change for different counterflow velocity $U$.
Finally, we supplement why and how we use the time-average method to describe the vortex number.

\section{1. Holographic superfluid model}

The holographic model provides a powerful tool for solving superfluid and other condensed matter systems, including equilibrium and non-equilibrium, through the duality of high-dimensional gravitational field and low-dimensional quantum field. Holographic dualities are now employed to study a wide range of non-equilibrium physical systems, including strongly coupled and dissipative superfluid. In our holographic two-component superfluid model, two complex scalar fleids coupled with two gauge fields are added to the bulk Einstein-Maxwell action, which is expressed as,
\begin{align}
S=\frac{1}{16\pi G_N} \int d^4x \sqrt{-g}\Big[\mathcal{L}_{EH}-\mathcal{L}_{matter}
\Big]
\\
\mathcal{L}_{EH}=R+6/L^2
\\
\mathcal{L}_{matter}=-\sum_{j=1}^2(\frac{1}{4}F^2_j-|D_j\Psi_j|^2-m_j^2|\Psi_j|^2)+\eta|\Psi_1|^2|\Psi_2|^2
\end{align}
where $\mathcal{L}_{EH}=R+6/L^2$ is the Einstein-Hilbert action with cosmological constant $6/L^2$ wiring in the asymptotic AdS geometry. $F_{j}=\partial_\alpha A_{j\beta}-\partial_\beta A_{j\alpha}$ is the Maxwell field strength with vector potential $A_{j\alpha}$, that is coupled minimally to the scalar involving the charge $q$ with the covariant derivative $D_j=\partial_{j\alpha}-iqA_{j\alpha}$.  $\Psi_j$ is the complex scalar field with mass  $m_j$. $\eta$ is the inter-component coupling constant.

In this context, we can adopt the convention of setting $m_j^2=-2$ without sacrificing the generality of our analysis. As per the holographic duality, the behavior of the bulk field near the boundary is characterized by $\Psi_j = \phi_j z + \psi_j z^2 + \mathcal{O}(z^3)$, while the gauge field $A_{j\alpha}({\bf r},z)$, which is dual to a conserved $U(1)$ current, exhibits asymptotic behavior given by $A_{j\alpha} =a_{j\alpha} +b_{j\alpha} z + \mathcal{O}(z^2)$.
Here, $\phi_j$ serves as a source term and is typically set to zero in the context of the spontaneously broken symmetry phase. On the other hand, $\psi_j$ corresponds to the vacuum expectation value $\langle O_j\rangle$ of the dual scalar operator, which exhibits a non-zero value in the broken phase. The quantities $a_{jx,y}$ and $b_{jx,y}$ represent the superfluid velocity and the associated conjugate current, respectively. Additionally, $a_{jt}$ is identified as the chemical potential $\mu$, while $b_{jt}$ denotes the charge density $\rho$ of the field theory.

The background metric employed in the Eddington coordinates takes the form of $ds^2=\frac{L}{z^2}[-f(z)dt^2-2dtdz+dx^2+dy^2]$, where $f(z) =1-(z/z_h)^3$, and the Hawking temperature is given by $T=3/(4\pi z_h)$. We consider a square boundary with periodic boundary conditions, where $x$ and $y$ denote the spatial coordinates, while $z$ represents the additional radial dimension within the bulk. Notably, the conformal invariance of the superfluid implies that its thermodynamics are exclusively governed by the only dimensionless parameter $\mu/T$.
Thus, when the chemical potential exceeds a critical value $\mu=4.07$, the temperature cooling to the broken phase.

When we disregard backreaction effects and consider the complete non-equilibrium, dynamic evolution system at the boundary, we can readily trace the evolution by numerically solving the remaining equations of motion within the bulk system. Within the bulk, the equilibrium geometry is described, and the equations of motion (EOM) governing the bulk gauge and scalar fields can be expressed as follows:
\begin{align}
d_\beta F^{\alpha\beta}_j = J^{\alpha}_j = iq(\Psi_j^*D_j^{\alpha}\Psi_j-\Psi_jD_j^{\alpha}\Psi_j^*) \\
(-D_j^2+m_j^2+\eta|\Psi_k|^2)\Psi_j=0
\end{align}
By imposing the holographic periodic boundary conditions, these equations can be solved numerically. In particular, we employed high-order Runge-Kutta methods  with time step $\Delta t = 0.01$. Additionally, the Chebyshev method was applied in the z-direction, while the Fourier method was utilized in the x and y directions when addressing the boundary conditions with the number of grid points taken as $20 \times 200 \times 200 (nz\times nx\times ny)$ and the boundary size taken as $40\times40 (Rx\times Ry)$.

We measure the number of vortices by calculating the phase winding number, expressed as $N = \frac{1}{2\pi} \oint \nabla\theta \cdot dr$, where $\theta$ is the phase of the wave function which is defined as $\langle O \rangle=\psi = |\psi| e^{i\theta}$. We identify the quantum vortices by using numerical method to judge the end point of phase where $\theta$ from $0$ to $2\pi$.
This approach ensures that multiply-charged vortices are adequately accounted for in our analysis.

\section{2. GROSS-PITAEVSKII EQUATION AND beyong-mean-field system}

 In the main text, we conduct simulations using the coupled Gross-Pitaevskii equations, here we introduce the Gross-Pitaevskii equation setup in detail, and then extend to the beyong-mean field case. The Gross-Pitaevskii lagrangian and equation we used in main text written as
\begin{align}
   \mathcal{L}_{GP}=\sum_{j=1}^2 \big[\frac{i\hbar}{2}(\Psi_j^*\frac{\partial\Psi_j}{\partial t}-\Psi_j\frac{\partial\Psi_j^*}{\partial t})-\frac{\hbar^2}{2m}|\nabla\Psi_j|^2+g_{jj}|\Psi_j|^4\big]+g_{12}|\Psi_1|^2|\Psi_2|^2\\
       i\hbar\frac{\partial}{\partial t}\Psi_j = (-\frac{\hbar^2}{2m_j}\nabla^2+\sum_kg_{jk}|\Psi_k|^2-{\bf u}_j \cdot {\bf p})\Psi_j
\end{align}

Here, we maintain consistency with the main text regarding the notation and parameters. The masses of both components are identical, denoted as $m_1 = m_2$ for the $j$-th component. The order parameter is represented as $\Psi_j = \sqrt{n_j} \exp(i\phi_j)$, corresponding to the complex scalar field or order parameter as used in the holographic superfluid context.

The coupling coefficient $g_{jk}$ signifies the atom interaction and is defined as $g_{jk} = 2\pi\hbar^2a_{jk}/m_{jk}$, with $m_{jk}^{-1} = m_j^{-1} + m_k^{-1}$ and the s-wave scattering length $a_{jk}$ between the $j$-th and $k$-th components. We consider $g_1=g_2$.  When the product of $g_{11}g_{22}>g_{12}^2$, the two components can be mixed effectively. Conversely, when the product of $g_{11}g_{22}<g_{12}^2$, the mixed state becomes unstable, leading to phase separation between the components.
The external flow term is ${\bf u}_j \cdot {\bf p}={\bf u}_j \cdot(-i\hbar\nabla)$, where ${\bf u}$ the flow velocity\cite{Reeves_2015}. Consistent with the text, we set a convective condition of periodic oscillation
\begin{equation}
     v_{jx}(t)=(-1)^{j-1}U \cos (\Omega t)
\end{equation}
 Similar to holographic simulations, we have employed Fourier methods to create two-dimensional periodic boundary conditions, with sizes of $40\times40(Rx\times Ry)$.

However, the Gross-Pitaevskii equation and holographic model, functioning as
a mean-field approximation, has its limitations, particularly concerning the conservation of
linear and angular momentum which is crucial to our discussion. So here we deal with a relatively convenient beyond-mean-field model
\begin{equation}
    i\hbar\frac{\partial}{\partial t}\Psi_j = (-\frac{\hbar^2}{2m_j}\nabla^2+\sum_kg_{jk}|\Psi_k|^2-{\bf u}_j \cdot {\bf p}-\frac{m^{1/2}g^{3/2}}{\pi\hbar}(|\Psi_i|^2+|\Psi_j|^2)^{1/2})\Psi_j
\end{equation}
The last term within the parentheses represents the beyond-mean-field correction due to quantum fluctuations from the Lee-Huang-Yang effect, thus, the equation encompasses higher-order many-body interactions.

\begin{figure}[h]
    \centering
    \includegraphics[width=12cm]{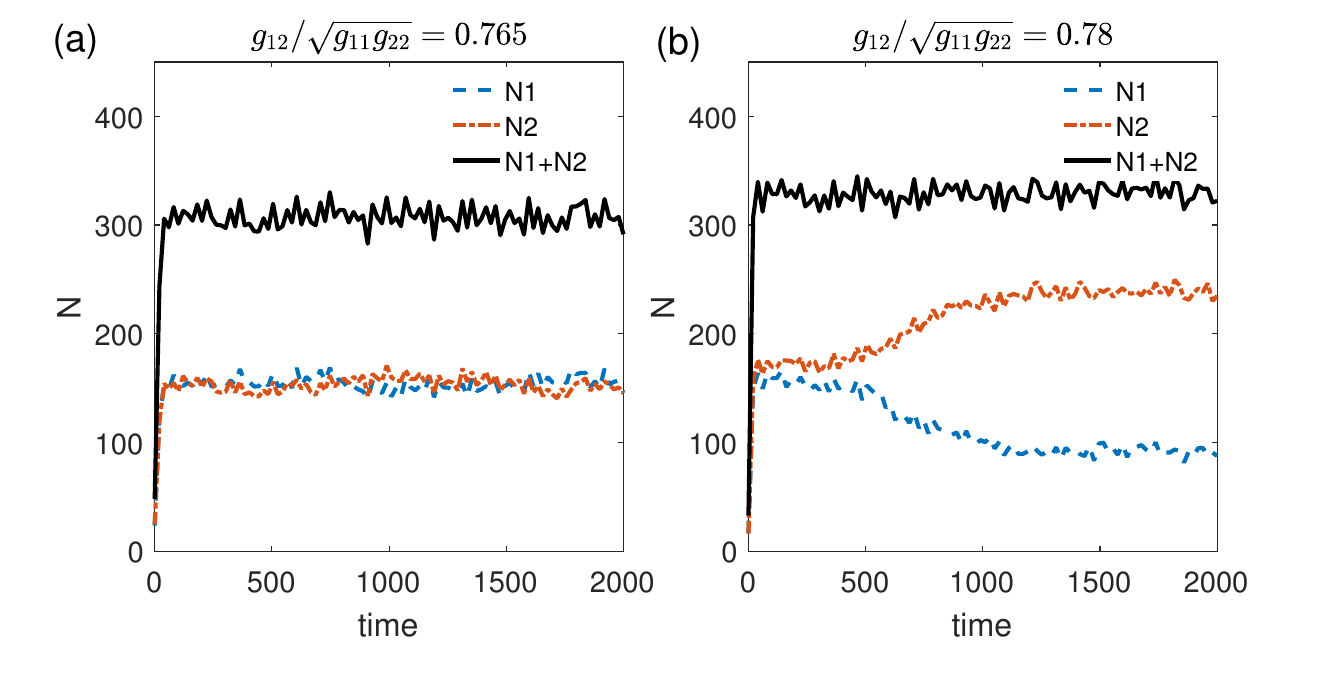}
    \caption{The evolution of vortex number as observed in beyond-mean-field simulations: {\bf (a):} The vortex number maintains symmetry. {\bf (b):} The vortex number exhibits symmetry breaking.}
    \label{figures1}
\end{figure}

As shown in Fig.\ref{figures1}, even in the beyond-mean-field systems, symmetry breaking in the number of vortices occurs in cases of high coupling constant. However, due to the additional coupling effects introduced by the beyond-mean-field, we anticipate changes in the phase diagram. This suggests that while the fundamental phenomenon of vortex number symmetry breaking is preserved, the specific dynamics and characteristics of the system may change under the influence of these extended fluctuations.
These additional results aim to demonstrate that the observations made in our study are not merely mean-field effects and could also be observed in a many-body system. 
\newpage

\section{3. Vortex symmetry breaking in non-periodic system}

All previous simulations were conducted in systems with periodic boundary conditions. However, 
in experimental settings, non-periodic systems with constrained potential fields are commonly employed.

\begin{figure}[h]
    \centering
    \includegraphics[width=12cm,trim= 0   0 0  0]{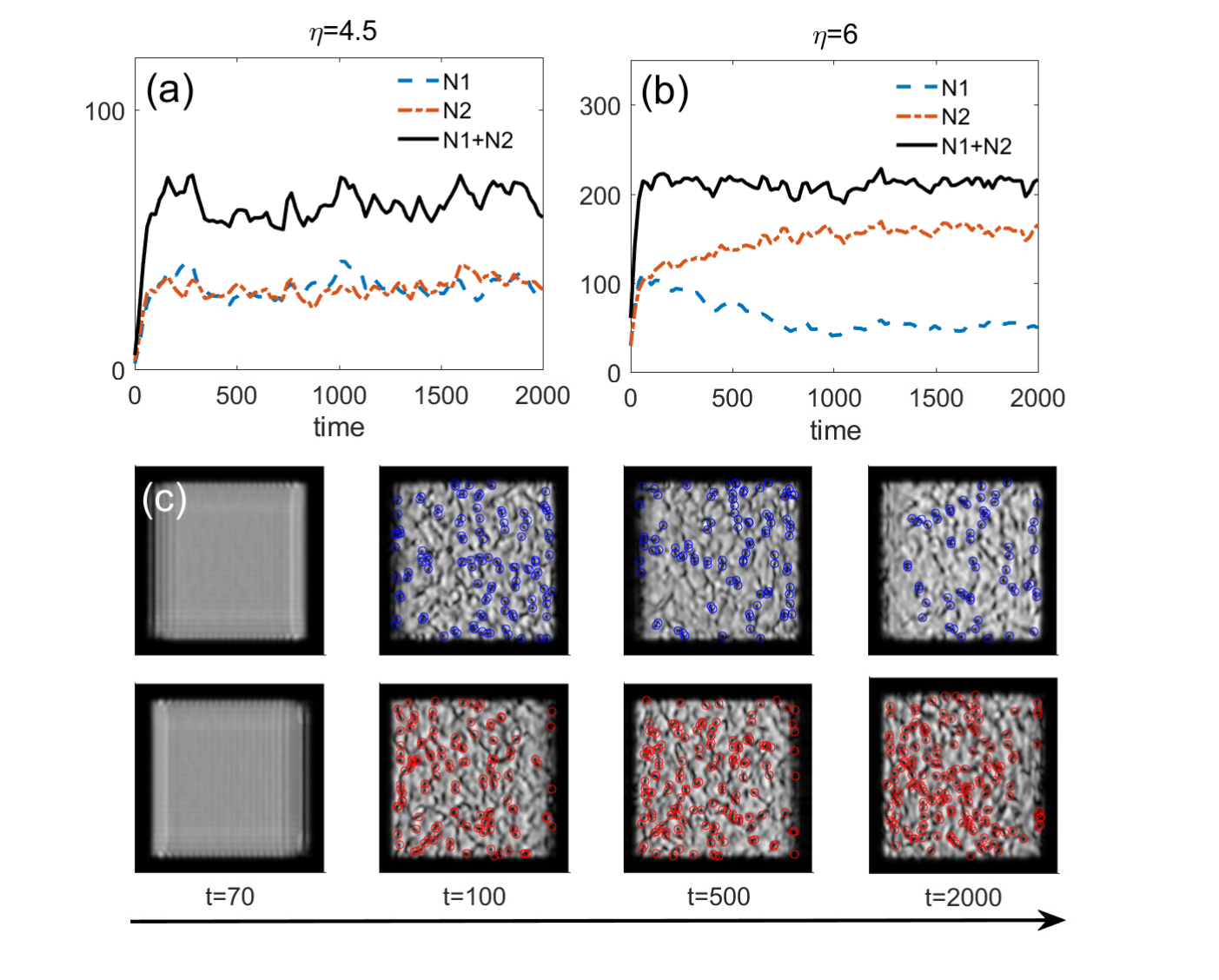}
    \caption{Figure of non-periodic system's simulation. {\bf (a):} The vortex evolution with coupling constant $\eta=4.5$, vortex number exhibits completely symmetrical behavior. {\bf (b):} The vortex evolution with coupling constant $\eta=6$, symmetry breaking occurs in the number of vortices.
    {\bf (c):} The dynamic evolution diagram of the densities corresponding to Figure (b). The upper and lower rows refer to components $\Psi_1$ and $\Psi_2$.}
    \label{figures2}
\end{figure}

In this section, in order to provide additional model examples and cater to experimental requirements, we conducted simulations using a non-periodic model with trapped potential barriers. 
Within a square boundary of size $40\times40$, we introduced square potential barriers of size $36\times36$, creating a non-periodic system. As indicated by the simulation results in Fig.\ref{figures2}, vortex number symmetry breaking is not limited to periodic systems but also occurs in commonly used non-periodic systems in experiments.

\section{4. Soliton structure and characteristic time}

In this section, we supplement our discussion with a more intuitive representation of the soliton's structure and phase. And we show the variation of the estimated characteristic time of soliton with the coupling coefficient.

\begin{figure}[h]
    \centering
    \includegraphics[width=8cm,trim=0 30 0 20]{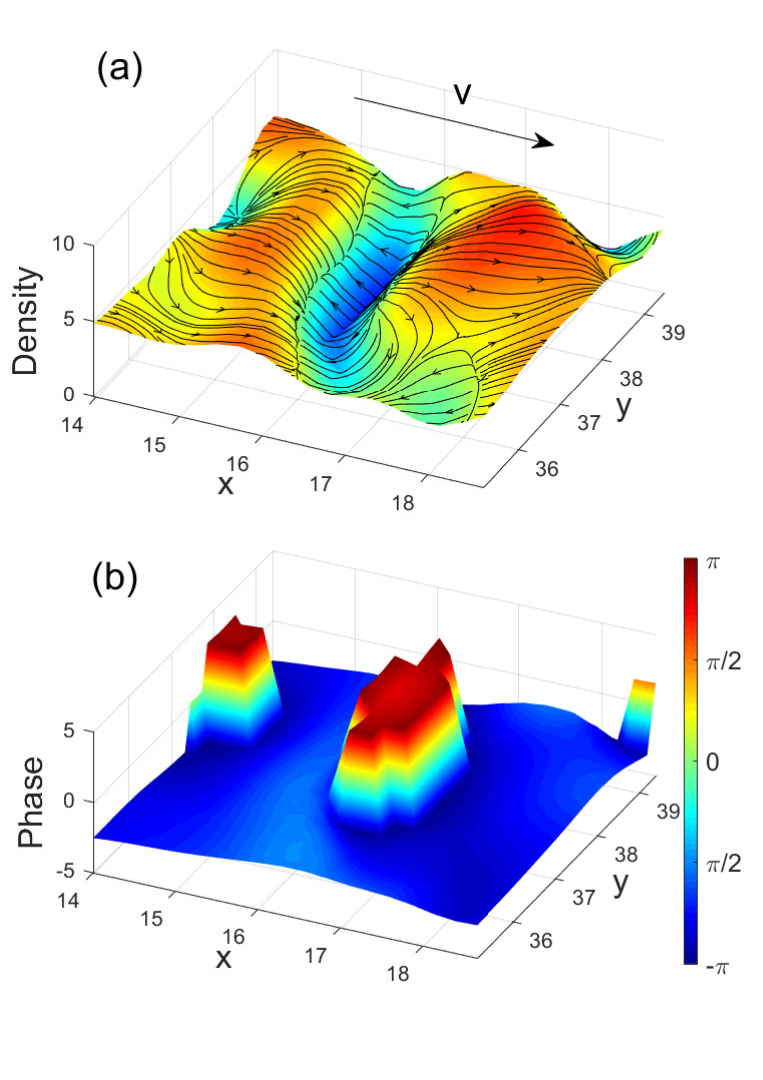}
    \caption{{\bf (a):} The density and velocity field of soliton structure. {\bf (b):} The phase of soliton.}
    \label{figures3}
\end{figure}

Fig.\ref{figures3}a illustrates the distribution of superfluid density and velocity field in the vicinity and interior of the soliton. It is understood that in the presence of countersuperflow instability, solitons initially form and eventually fragment into vortex pairs due to this instability. However, in the Alternating Current (AC) case, if the direction of the background superflow changes before the soliton divides, the flow velocity within the soliton differs from that outside, as depicted in the figure. This discrepancy leads to an alteration in the amplitude of the average velocity when there is asymmetry in the number of vortices.

Fig.\ref{figures3}b depicts the phase distribution associated with the soliton, highlighting a significant phase drop of approximately $\pi$ at the edge of the soliton. 

The lifetime of solitons varies individually, which can be observed from the fact that in DC counterflow, vortices are not generated instantaneously. Therefore, in the maintext, we consider the characteristic times of soliton splitting, or average lifetimes. 

In AC counterflow, when solitons cannot completely split within half a period,  differences in flow velocity will appear in the next half period, thereby causing an imbalance. 

\begin{figure}[h]
    \centering
    \includegraphics[width=9cm,trim= 0 0 0 0]{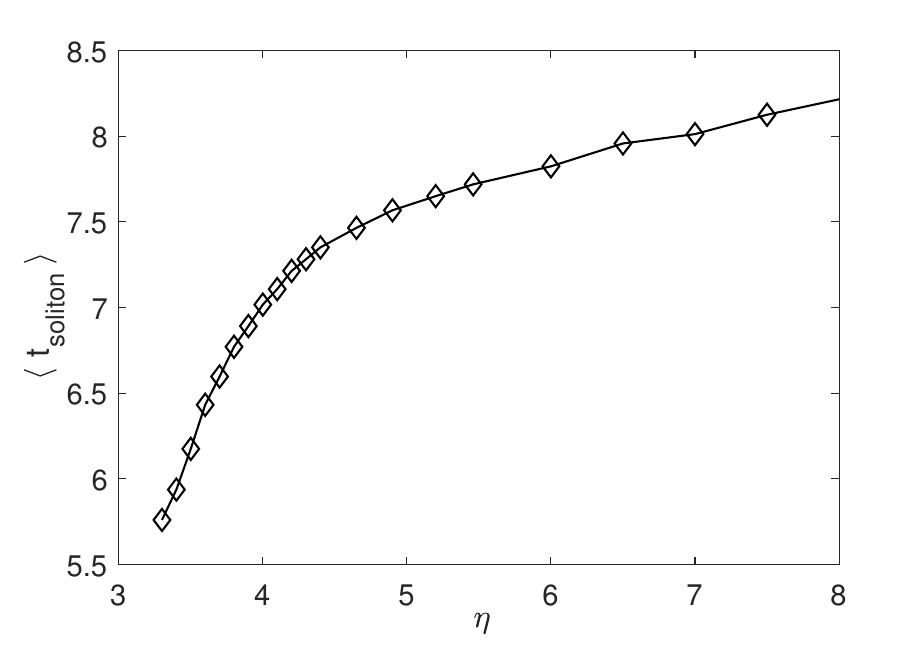}
    \caption{Variation of the characteristic time for soliton splitting as a function of the coupling constant.}
    \label{figures4}
\end{figure}

Hence, the characteristic time of solitons can be estimated from the critical frequency. $\langle t_{soliton} \rangle= \pi/\Omega_c$. This value will change with the alterations in the coupling constant. As shown in Fig.\ref{figures4}, the characteristic time of soliton increases with the increase of coupling constant.

\section{5. DC counterflow with equal and unequal velocities}

As a comparative analysis with AC counterflow, and to illustrate the effect of flow velocity on vortex nucleation, we present the results of DC counterflow here.
 Contrary to AC counterflow, where vortices are continuously generated and annihilated, in DC counterflow, 
vortex will nucleate initially because the disturbance as a seed and
then decay to the equilibrium state that satisfies the solution of the equation of motion, eventually leaving only irrotational fluid in the system. Therefore, differing from AC case, in DC counterflow, all solitons will split into vortices even at sufficiently high velocities U. This results in no imbalance due to the symmetry of the
two components.

\begin{figure}[h]
    \centering
    \includegraphics[width=9cm]{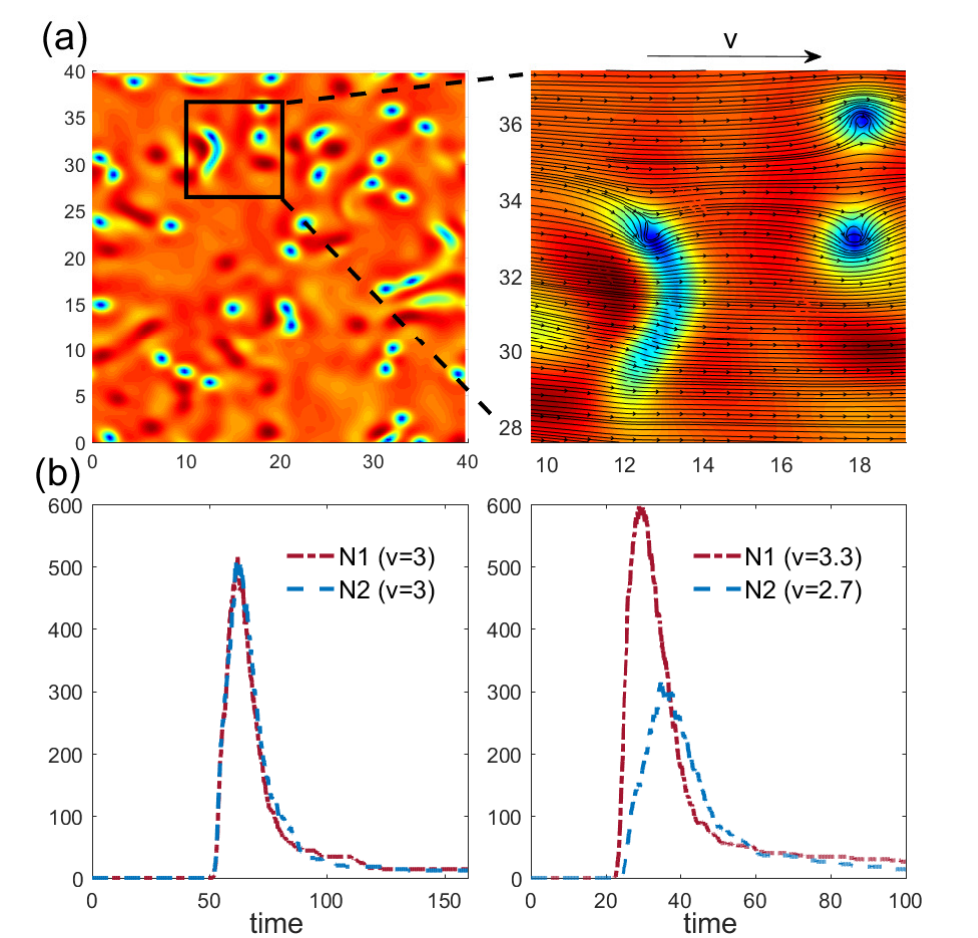}
    \caption{ {\bf (a)}: Global density distribution diagram and local enlarged velocity field diagram of soliton and vortex in DC counterflow. {\bf (b)}: Left panel shows the vortex number in counterflow with equal velocity of two components. Right panel shows the vortex number of two components with different velocity.}
    \label{figures5}
\end{figure}

In Fig.\ref{figures5}a, 
we observe that the actual superflow velocity remains entirely consistent with the background velocity, even in the presence of a Soliton-like structure. This consistency is due to the absence of background velocity switching, which in turn ensures that the soliton's flow velocity remains consistently aligned with the background flow. This fundamental distinction underlies the difference between AC counterflow and DC counterflow.

In the left panel of Fig.\ref{figures5}b, we observe that when the two components of the counterflow share the same velocity, the vortex count is equal for both. However, the right panel reveals that when these components possess differing velocities, the slower component exhibits a reduced rate of vortex generation. This observation elucidating the expansion of symmetry breaking in AC counterflow when solitons are introduced.

\section{6. Total vortex number versus coupling constant $\eta$}

\begin{figure}[h]
    \centering
    \includegraphics[width=9.3cm]{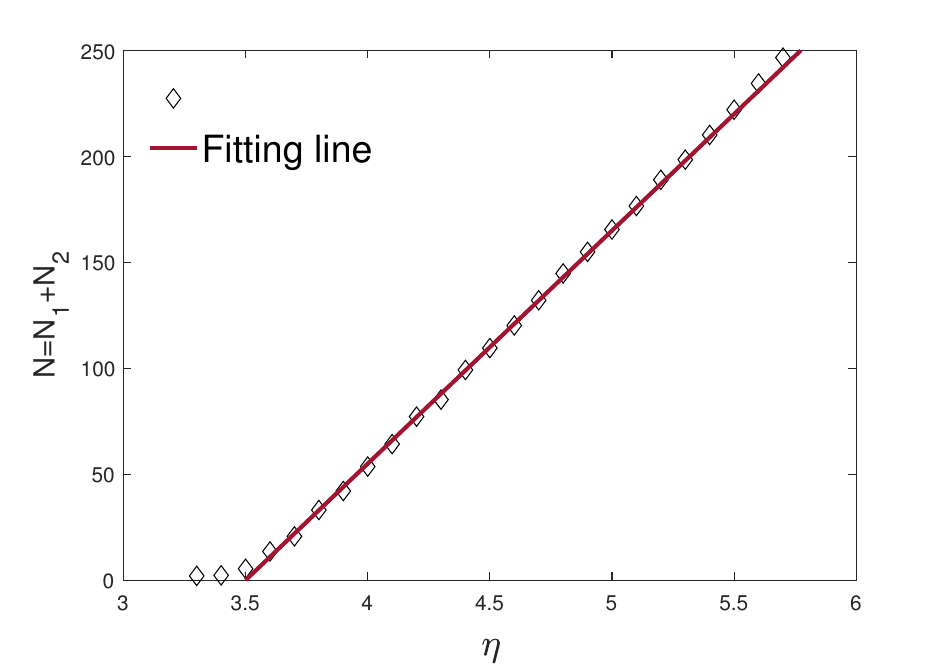}
    \caption{The total vortex number versus coupling coefficient. Black diamond dot is the numerical result, red line is the fitting curve as $N=(114\pm1.6)\eta-402.4\pm7.8$.}
    \label{figures6}
\end{figure}

In line with the instability conditions linked to counterflow, vortex nucleation is initiated when the counterflow velocity and the coupling coefficient surpass their respective critical thresholds. As illustrated in Fig.\ref{figures6}, beyond this critical coupling coefficient, there is a clear proportional relationship between the total number of vortices and the magnitude of the coupling coefficient. This direct proportionality highlights the sensitivity of vortex generation to changes in the coupling strength within the system.\newpage

\section{7.Phase diagram with different counterflow velocity}

\begin{figure}[h]
    \centering
    \includegraphics[width=13cm,trim= -50 50 0 0]{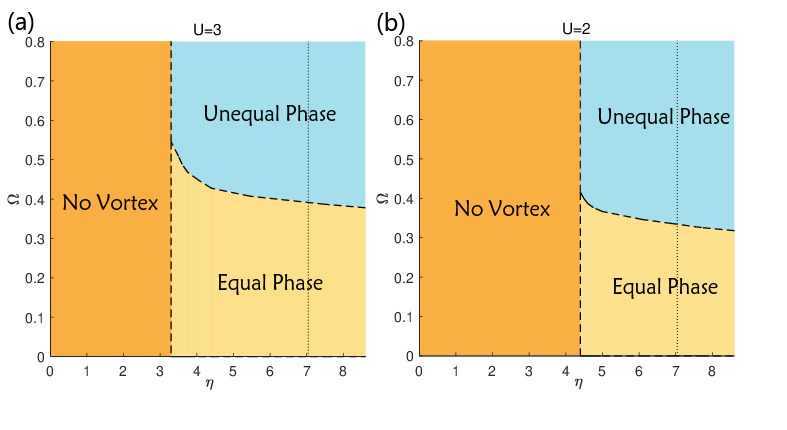}
    \caption{{\bf (a):} The phase diagram as detailed in the main text. {\bf (b):} An altered phase diagram with counterflow velocity $U=2$.}
    \label{figures7}
\end{figure}

In this section, we demonstrate how the phase diagram changes with the counterflow velocity. 
As shown in Fig.\ref{figures7}, the left image is the phase diagram as described in the main text, with a counterflow velocity of $U=3$, while the right image is a newly generated phase diagram for $U=2$. 
It can be observed that when the counterflow velocity decreases, several changes occur. Firstly, the critical coupling constant for vortex nucleation increases. This observation aligns with our theory, as both the counterflow velocity and the coupling coefficient are positively correlated with the number of vortices, so the reduction of the counterflow velocity means that greater coupling is needed to nucleate the vortices.
 Additionally, we observe that for the same coupling constant, the critical frequency for symmetry breaking decreases, because a lower counterflow velocity implies a longer characteristic time for soliton splitting, resulting in a lower critical frequency according.

\section{8.Time-average method used to deal with vortex number}

\begin{figure}[h]
    \centering
    \includegraphics[width=13cm,trim= 0   0 0  0]{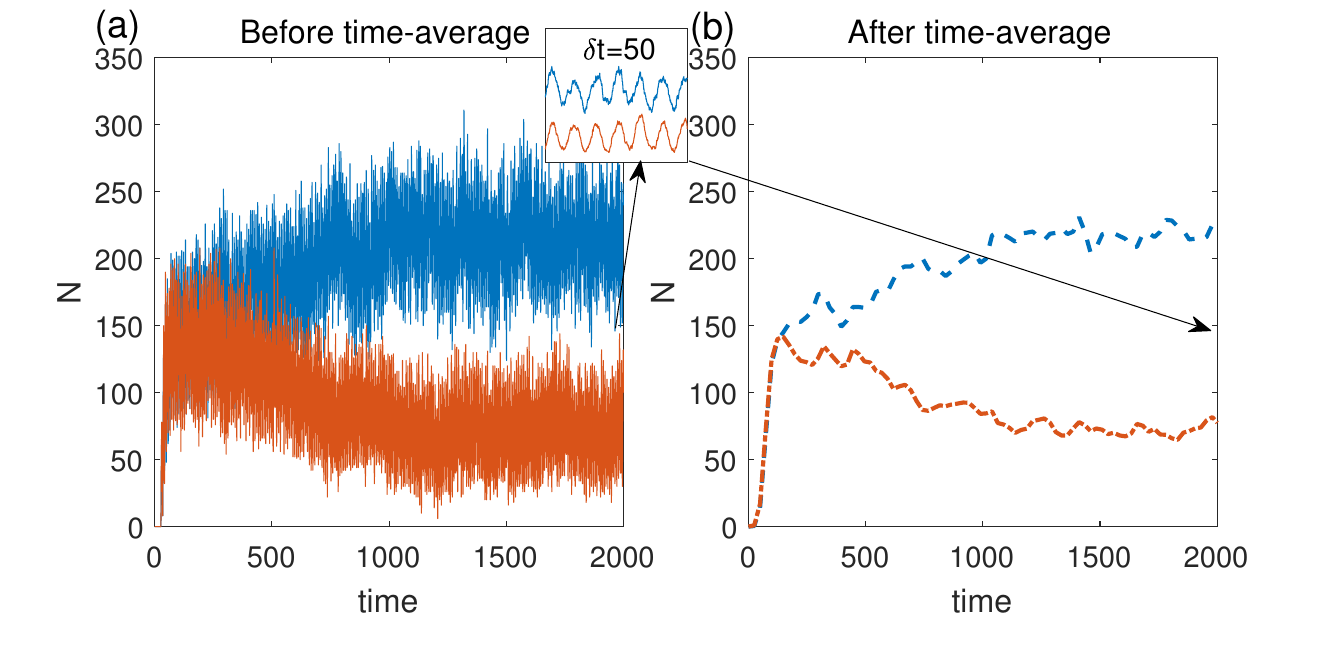}
    \caption{Comparison of vortex number before and after time-averaging treatment, with the left image showing before treatment and the right image showing after. The inset shows a magnified view of the vortex changes in the treatment interval $\delta t$.}
    \label{figures8}
\end{figure}

In the main text, we employed a time-averaging method to process the vortex number. This was necessary because, due to the driving force of the periodic potential, the vortex number also varies periodically, as shown in Fig.\ref{figures8}a. In order to clearly distinguish the results of vortex number symmetry breaking, we performed a time-averaging of the vortex number for each point by considering a neighborhood within a time interval of $\delta t=50$. This results in the smooth curve shown in Fig. \ref{figures8}b.

For the determination of the final vortex number for each set of parameters, we similarly selected the time average over a long-duration region of the non-equilibrium steady state, representing the vortex number at the end of the simulation.

\end{document}